\documentclass[preprintnumbers,article,amsmath,amssymb,floatfix,10pt,prd,onecolumn,
superscriptaddress,nofootinbib]{revtex4}
\usepackage[colorlinks=true, pdfstartview=FitV, linkcolor=blue, citecolor=red, urlcolor=magenta]{hyperref}
\usepackage{bbm}
\usepackage{amsfonts}
\usepackage{mathrsfs}
\usepackage{latexsym}
\usepackage{epsfig}
\usepackage{epstopdf}
\usepackage{epstopdf}
\usepackage{graphicx}
\usepackage{amssymb}
\usepackage{amsmath}
\usepackage{dcolumn}
\usepackage{bm}
\usepackage{float}
\usepackage{color}
\usepackage{comment}
\usepackage{xcolor}
\begin{document}
\title{Gray-body factor and absorption of the Dirac field in ESTGB gravity}
\author{Qian Li}
\affiliation{Faculty of Science, Kunming University of Science and Technology, Kunming, Yunnan 650500, China.}
\author{Chen Ma}
\affiliation{Faculty of Science, Kunming University of Science and Technology, Kunming, Yunnan 650500, China.}
\author{Yu Zhang}
\email{zhangyu\_128@126.com Corresponding author} \affiliation{Faculty of Science, Kunming University of Science and Technology, Kunming, Yunnan 650500, China.}
\author{Zhi-Wen Lin}
\affiliation{Faculty of Science, Kunming University of Science and Technology, Kunming, Yunnan 650500, China.}
\author{Peng-Fei Duan}
\affiliation{City College, Kunming University of Science and Technology, Kunming, Yunnan 650051, China.}
\begin{abstract}
The gray-body factor and the absorption cross section of the 4D ESTGB gravity with a mode of nonlinear electrodynamics for the  massless Dirac field are studied in this paper. The magnetic charge value varies between $-2^{(\frac{5}{3})}/3$ and $0$ as well as  the ADM mass is set to $1$, which corresponds to a non-extreme black hole. The gray-body factor is obtained using the semi-analytic WKB method after solving the massless Dirac equation. When the absolute value of  magnetic charge is increasing, the gray-body factor $\gamma(\omega)$ is decreasing. In addition, the partial absorption cross section and the total absorption cross section are calculated by using the partial wave method. We find that the maximum value of partial absorption cross section decreases as $\kappa$ increases. And the existence of magnetic charge causes the diminishing of the total absorption cross section. Finally, we find  that  the absorption cross section of the  Dirac field is more sensitive to  electric charge than magnetic charge by comparing the absorption cross section of the  Reissner-Nordstr$\rm\ddot{o}$m  and ESTGB-NLED black holes.
\end{abstract}
\date{\today}
\maketitle
\section{Introduction}

In the decades after Einstein's general theory of relativity predicted the existence of black holes, the research on the related properties of black holes has gradually become a frontier hot issue in astrophysics. An important feature of black holes is that no information can escape from its event horizon. As a result, the existence of black holes can only be detected by a lot of the indirect methods in the previous decades. Nevertheless, with the development of technology, the Event Horizon Telescope Cooperation Organization has successfully captured the first image of a black hole in the center of the M87 galaxy \cite{1}. This picture directly proves the existence of black holes in the universe. However, a lot of phenomenons cannot be explained by the basic general relativity. These phenomenons include but are not limited to the acceleration expansion stage of the universe \cite{Riess1998}, the combination of gravity and the laws of quantum physics \cite {Fradkin1985}, and the flatness of the spiral curve for a spiral galaxy \cite{Capozziello2007}. Accordingly, this presents a wide room for altering and interpreting theories of gravity, such as the so called extended scalar-tensor-Gauss-Bonnet(ESTGB) \cite{Doneva2018}, which is a theory in four dimensions. Specifically, the scalar field is coupled with Gauss-Bonnet invariant in ESTGB to avoid the Ostrogradsky instability \cite{Woodard02210}. The black hole solutions have been presented by solving the complex field equation in ESTGB gravity without matter field in four dimensions. In addition, these numerical solutions are also given in the context of different matter fields, for instance, massive scalar \cite{Doneva2019}, charged case \cite{Doneva2018},  Dilatonic \cite{Kanti1996},  multi-scalar\cite{scalariz_Doneva}, and a particular form of nonlinear electrodynamics \cite{Canate2020}.

Black hole that is not an isolated system will interact with the surrounding environment. Interactions present more interesting phenomena, such as radiation, absorption, and scattering. Therefore, we can study how black holes interact with the surrounding environment to obtain interrelated information about special objects. In addition, the experiments to explore black holes rely largely on GW astronomy, shadow images and X-ray spectroscopy. In a way, all three aspects depend on the effect of black holes on the environment. As we all know, the accretion plays a non-negligible role in the  phenomenology  of active galactic nuclei \cite{Macedo2014}. Accretion of fundamental field, i.e., scalar field, electromagnetic field and Dirac field etc., is usually associated with the research of  absorption cross section. So it is necessary to research  the absorption  of waves and particles by black hole. Since the 1960s, theorists have begun to study the  problems related to scattering. Moreover, the gray-body  factor can help us understand the absorption and scattering of particles, and it is also an important factor to solve Hawking radiation. Hawking radiation \cite{Hawking1976} proposed by  Hawking in 1976, which depends on the gray-body factor and black hole temperature, is of crucial importance when studying the black hole information paradox. It may be the most difficult obstacle to a thorough understanding of quantum gravity.  The gray-body factor is defined as the possibility  that the incident particle with frequency $\omega$ is absorbed by the black hole, which  encodes valuable information about the near-horizon structure and correlative physics of the black hole. And it is used to measure the deviation from the radiation of the ideal and perfect black-body \cite{Kanti2002}.  Many authors \cite{Songbai2010,Ama-Tul-Mughani2021,Sharif2020,Qanitah Ama-Tul-Mughani2020,Konoplya2021} have also studied the gray-body of various black holes with the different methods.
	
A plethora of methods have been proposed to calculate  the gray-body factor with different accuracy, including  the new cancellation between contributions to the wave function for the different spin particles \cite{Cvetic1998},  the exact numerical method \cite{Zhang2018}, the rigorous  bounds for the  gray-body factor \cite{Miao2017}, the WKB method\cite{Konoplya2011}, etc. The WKB approximation stands out among the above-mentioned various calculation methods because of its versatility and flexibility. Blome, Hans-Joachim and Mashhoon \cite{Blome1984} proposed the first simple semi-analytic formula to calculate the quasinormal frequency by matching the effective potential with the inverse P$\rm\ddot{o}$schl-Teller potential. However, this formula fails to improve the accuracy for the lower multi-pole numbers. A year later, Schutz and Will \cite{Schutz1985} calculated the quasinormal modes using the WKB approximation based on Mashhoon's formula. The method is to match the WKB solution with the Taylor expansion that passes two turning points. Subsequently, Iyer, Sai and Will \cite{Iyer1987} introduced the third-order formula of the WKB method, which improved the accuracy up to one percent on the basis of Schutz and Will. Moreover, Konoplya \cite{Konoplya2003} and  Matyjasek et al. \cite{Matyjasek2017} proposed higher WKB order terms.  The WKB approximation can calculate not only the gray-body factor, but also the quasinormal mode. The quasinormal modes \cite{Wongjun2020,Cai2020,Hu2020,Liang2018,Saleh2018,Liang20018,Aragon2021,Li2017,Wang2021,Jawad2020,Sharif2021,Guo2013} containing complex frequencies represent the response of black holes to external perturbations such as massless scalar fields, neutrino fields, gravitational fields, electromagnetic fields, Dirac fields, etc.

In the 1970s, Hawking discovered that  the evaporation rate of a black hole is directly proportional to its  absorption cross section \cite{Futterman}. Subsequently, a wealth of important researches on the absorption and scattering of plane waves acting on black holes were established in the 1970s and 1980s. For instance, Sanchez \cite{Sanchez1978} indicated that the absorption cross section of a Schwarzschild black hole for the  massless scalar field is oscillatory with respect to the geometry-optical limit $(27/4)\pi r_{s}^2$, and Unruh \cite{Unruh1976} studied the absorption in the massive scalar. Besides, Crispino \cite{Crispino2007} presented the absorption of electromagnetic waves in the  Schwarzschild spacetime for arbitrary frequencies,  and Jung \cite{Jung2004} studied the absorption of massive scalar field for the Reissner-Nordstr$\rm\ddot{o}$m spacetime. Next, the result of absorption of electromagnetic waves was obtained in Ref. \cite{Crispino2008}. In addition, the absorption of massless scalar by kerr  spacetime was investigated in Ref. \cite{Macedo2013}, and the absorption of electromagnetic waves was analyzed in  Ref. \cite{Leite2017}. Liao Hao, Songbai Chen et al. \cite{Songbai Chen2014} analyzed the absorption and Hawking radiation of electromagnetic waves with Weyl correction in 4D black hole spacetime. A lots of authors \cite{Huang2019,Huang2015,2018, Anacleto2020,Magalhaes2020,Junior2020,Benone2018} have also studied the absorption and scattering cross sections of various black holes. In this paper we will study the gray-body factor and absorption cross section of the black hole in ESTGB gravity with an unusual form of nonlinear electrodynamics in four dimensions.
	
This paper is organized as follows. The second section outlines the basic information of the four-dimensional (4D) extended scalar-tensor-Gauss-Bonnet theory (ESTGB) coupled with a special form of nonlinear electrodynamics, and the settings of related parameters are also given. In the third section, the massless Dirac equation is reduced to master wave equations and the effective potential is analyzed. Next, the gray-body factor is calculated using the WKB method in the fourth section. The fifth section presents the expression of the absorption cross section of the  Dirac field and the corresponding  results are also given. Summary and conclusions are presented in the last section.

\section{The black hole solution in the extended scalar-tensor-Gauss-Bonnet gravity}\label{sec:2}

Without loss of generality, we adopt natural units in this paper, namely c = G = $\hbar$ = 1. The $4D$ extended scalar-tensor-Gauss-Bonnet theory coupled with a particular form of nonlinear electrodynamics (ESTGB-NLED) \cite{Canate2020} is defined as follows,
\begin{eqnarray}\label{action}
	S = \int d^{4}x \sqrt{-g} \left\{ \frac{1}{16\pi}\left(R - \frac{1}{2}\partial_{\mu}\phi\partial^{\mu}\phi + \boldsymbol{f}(\phi) R_{_{GB}}^{2} - 2 \cal U (\phi) \right)
- \frac{1}{4\pi}\mathcal{L}_{\rm matter}  \right\}.
	\end{eqnarray}
	where  $R$ is the Ricci scalar, $\phi$ is the scalar field, $\boldsymbol{f}(\phi)$ is a coupling function that depends only on $\phi$, $R_{_{GB}}^{2}$ is the Gauss-Bonnet term and $ \cal U (\phi)$ means the scalar field potential. The first term is the Einstein-Hilbert Lagrangian density. The Lagrangian density $\mathcal{L}_{\rm matter}$  represents any matter field. Assuming that the metric is static, then we have the following spherically symmetric form,
\begin{eqnarray}\label{1}
	\text{d}s^2=-f(r)\text{d}t^2+f^{-1}(r)\text{d}r^2+r^2\text{d}\Omega^2,
	\end{eqnarray}
with
\begin{eqnarray}
	f(r)=1-\frac{2m}{r}-\frac{q^3}{r^3},
\end{eqnarray}
	where $m$ represents the ADM mass and $q$ indicates the magnetic charge. Supposing that both the effective energy-momentum tensor and the corresponding NLED energy-momentum tensor satisfy the weak energy condition, we can obtain that $m>0$ and $q<0$. The  metric is similar to those used in the Reissner-Nordstr$\rm\ddot{o}$m black hole that retains two horizons, one horizon, or none. Moreover, the Schwarzschild black hole is recovered when $q$ is set to $0$. Without loss of generality, we  consider the black hole is non-extreme \cite{Canate2020} satisfying the value of $ 0 > q/m > -2^{(\frac{5}{3})}/3$. If we set the ADM mass $m=1$ and then the magnetic charge $-2^{(\frac{5}{3})}/3<q<0$.

For  later  comparisons, it is beneficial to explicitly mention the Reissner-Nordstr$\rm\ddot{o}$m black hole solution. The metric of Reissner-Nordstr$\rm\ddot{o}$m black hole  is given by

\begin{eqnarray}
	\text{d}s^2=-(1-\frac{2M}{r}+\frac{Q^2}{r^2})\text{d}t^2+(1-\frac{2M}{r}+\frac{Q^2}{r^2})^{-1}\text{d}r^2+r^2\text{d}\Omega^2,
\end{eqnarray}
where $Q$ is the electric charge of the black hole and $M$ is the mass of black hole.

\section{Master wave equation in Dirac field}\label{sec:3}

	In this section, we reduce the massless Dirac equation in the black hole spacetime to a series of Schr$\rm\ddot{o}$dinger-like radial equations, and analyze the properties of the effective potential. The massless Dirac equation in black hole spacetime has the following form  according to Ref. \cite{Brill1957},
	\begin{equation}\label{covdirac}
	\gamma^{\alpha} \ e_{\alpha}^{\mu} \left( \partial_{\mu} + \Gamma_{\mu} \right) \Psi=~0,
	\end{equation}
	where $\gamma^{\alpha}$ mean the Dirac matrices, defined as follows,
	\begin{equation}\label{covdirac2}
	\gamma^{0}=\left(
	\begin{array}{c c}
	-i & 0  \\
	0 & i
	\end{array}
	\right),
	~~\gamma^{i}=\left(
	\begin{array}{c c}
	0 & -i\sigma^{i} \\
	i\sigma^{i} & 0
	\end{array}
	\right),
	\end{equation}
	with $\sigma_{i} $ representing the pauli matrix,  for any $i \in \{1,2,3\}$, respectively.
	Moreover, $ e_{\alpha}^{\mu}$ is the inverse of the tetrad  $ e_{\mu}^{\alpha}$, of which the particular form is  defined by the metric $g_{\mu\nu}$,
	\begin{equation}\label{gmetric}
	g_{\mu\nu}=\eta_{ab} e_{\mu}^{a} e^{b}_{\nu},
	\end{equation}
	where $\eta_{ab}$ is the Minkowski metric, and $\eta_{ab}=\text{diag}(-1,1,1,1)$. Additionally, $\Gamma_{\mu}$ is the spin connection defined by
	\begin{equation}\label{tau}
	\Gamma_{\mu}=\frac{1}{8}[\gamma^{a},\gamma^{b}] e_{a}^{\nu} e_{b\nu;\mu},
	\end{equation}
	with $e_{b\nu;\mu}=\partial_{\mu} e_{b\nu}-\Gamma_{\mu\nu}^{\alpha} e_{b\alpha}$ being the covariant derivative of $e_{b\nu}$, where $\Gamma_{\mu\nu}^{\alpha}$ is the Christoffel symbol.
	In the spacetime of a static and spherical black hole, $e_{\mu}^{a}$ can be expressed as
	\begin{equation}\label{diag}
	e_{\mu}^{a}=\text{diag}(\sqrt{f},\frac{1}{\sqrt{f}},r,r\sin\theta).
	\end{equation}
	Hence, the components of $\Gamma_{\mu}$ can be obtained by substituting equation (\ref{diag}) into equation (\ref{tau}) as follows,
	\begin{equation}\label{GAMMA}
	\Gamma_{0}=\frac{1}{4} f^{'} \gamma^{1} \gamma^{0},\Gamma_{1}=0,\Gamma_{2}=\frac{1}{2}\sqrt{f}\gamma^{1}\gamma^{2},\Gamma_{3}=\frac{1}{2}(\sin\theta\sqrt{f}\gamma^{1}\gamma^{3}+\cos\theta\gamma^{2}\gamma^{3}).
	\end{equation}
	
	Further substituting the above expressions into the Dirac equation (\ref{covdirac}), the Dirac equation becomes
	\begin{equation}\label{gamma}
	\frac{\gamma^{0}}{\sqrt{f}}\frac{\partial{\Psi}}{\partial{t}}+\sqrt{f}\gamma^{1}(\frac{\partial}{\partial{r}}+\frac{1}{r}+\frac{1}{4f}\frac{ df}{ dr})\Psi+\frac{\gamma^{2}}{r}(\frac{\partial}{\partial\theta}+\frac{1}{2}\cot\theta)\Psi+\frac{\gamma^{3}}{r\sin\theta}\frac{\partial{\Psi}}{\partial{\psi}}=~0.
	\end{equation}
	Furthermore, we can transform equation (\ref{gamma}) into the following equation
	\begin{equation}\label{GAMMA}
	\frac{\gamma^{0}}{\sqrt{f}}\frac{\partial{\Phi}}{\partial{t}}+\sqrt{f}\gamma^{1}(\frac{\partial}{\partial{r}}+\frac{1}{r})\Phi+\frac{\gamma^{2}}{r}(\frac{\partial}{\partial\theta}+\frac{1}{2}\cot\theta)\Phi+\frac{\gamma^{3}}{r\sin\theta}\frac{\partial{\Phi}}{\partial{\phi}}=~0,
	\end{equation}
	by defining a tortoise coordinate change as
	\begin{eqnarray}
	r_{\star}=\int \frac{dr}{f},
	\end{eqnarray}
	introducing the ansatz  as
	\begin{eqnarray}
	\Phi=\left(
	\begin{array}{c}
	\frac{i G^{\pm}(r)}{r}\phi_{jm}^{\pm}(\theta,\varphi) \\ \frac{F^{\pm}(r)}{r}\phi_{jm}^{\mp}(\theta,\varphi)
	\end{array}
	\right) e^{-iwt},
	\end{eqnarray}
	and defining spinor angular harmonics as
	\begin{eqnarray}
	\phi_{jm}^{+}=\left(
	\begin{array}{c}
	\sqrt{\frac{l+\frac{1}{2}+m}{2l+1}}Y_{l}^{m-\frac{1}{2}} \\ \sqrt{\frac{l+\frac{1}{2}-m}{2l+1}}Y_{l}^{m+\frac{1}{2}}
	\end{array}
	\right),     (j=l+\frac{1}{2}),
	\end{eqnarray}
	\begin{eqnarray}
	\phi_{jm}^{-}=\left(
	\begin{array}{c}
	\sqrt{\frac{l+\frac{1}{2}-m}{2l-1}}Y_{l}^{m-\frac{1}{2}} \\ -\sqrt{\frac{l+\frac{1}{2}+m}{2l-1}}Y_{l}^{m+\frac{1}{2}}
	\end{array}
	\right),     (j=l-\frac{1}{2}).
	\end{eqnarray}
	
	As a result, equation (\ref{gamma}) can be rewritten as
	\begin{eqnarray}\label{Rewritten16}
	\left(
	\begin{array}{cc}
	0 & -\omega  \\
	\omega & 0
	\end{array}
	\right)
	\left(
	\begin{array}{c}
	F^{\pm} \\
	G^{\pm}
	\end{array}
	\right)-\frac{\partial}{\partial{r_{\star}}} \left(
	\begin{array}{c}
	F^{\pm} \\
	G^{\pm}
	\end{array}
	\right)+\sqrt{f}
	\left(
	\begin{array}{cc}
	\frac{k_{\pm}}{r} & 0 \\
	0 & -\frac{k_{\pm}}{r}
	\end{array}
	\right)
	\left(
	\begin{array}{c}
	F^{\pm} \\
	G^{\pm}
	\end{array}
	\right)=~0,
	\end{eqnarray}
	where the different cases for $(+)$ and $(-)$ in the function $F^{\pm} $ and $G^{\pm} $ are given by \cite{Cho2005}
	\begin{eqnarray}\label{Rewritten17}
	\frac{d^2{F}}{d{r_{\star}^{2}}}+(\omega^{2}-V_{1})F=~0, \\
	\frac{d^2{G}}{d{r_{\star}^{2}}}+(\omega^{2}-V_{2})G=~0,
	\end{eqnarray}
	with
	\begin{eqnarray}
	V_{1} = \frac{\sqrt{f}\left|\kappa\right|}{r^{2}}(\left|\kappa\right|\sqrt{f}+\frac{r}{2}\frac{df}{dr}-f),(for~ \kappa=j+\frac{1}{2},~~~~~and ~j=l+\frac{1}{2}),  \\
	V_{2} = \frac{\sqrt{f}\left|\kappa\right|}{r^{2}}(\left|\kappa\right|\sqrt{f}-\frac{r}{2}\frac{df}{dr}+f),(for~ \kappa=-(j+\frac{1}{2}),~and~ j=l-\frac{1}{2}).
	\end{eqnarray}
	
	It is worth noting that the potentials $V_{1}$ and $V_{2}$ are super-symmetric  partners  \cite{Cooper1995}, and they are derived from the same super-potential. It is well established that the potentials $V_{1}$ and $V_{2}$  related in this way have the same spectra. Therefore, we only need to consider the effective potential $V_{1}$ in calculating the gray-body factor and the absorption cross section for the  massless Dirac field by the WKB approximation. As a result, the equation (\ref{Rewritten17}) can be written as
	\begin{eqnarray}\label{E_nled2}
	\frac{d^2{\psi}}{dr_{\star}^{2}}+(\omega^{2}-V_{eff})\psi=0.
	\end{eqnarray}

\begin{figure}[htbp]
	\centering
	{
		\begin{minipage}[t]{0.5\linewidth}
			\centering
			\includegraphics[width=3.3in]{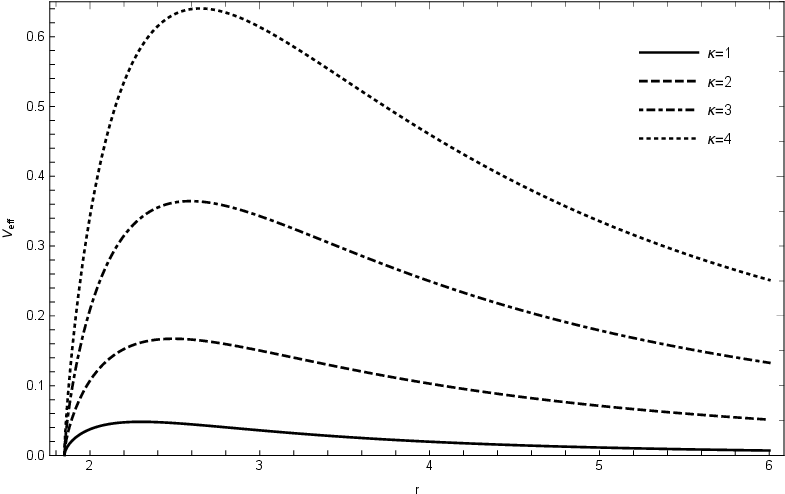}
		\end{minipage}%
	}%
		\caption{Behavior of the effective potential with different values of $\kappa$ for $q=-0.8$.}
\label{eff2}
\end{figure}

\begin{figure}[htbp]
\begin{minipage}[t]{1\linewidth}
\centering
\includegraphics[width=3.3in]{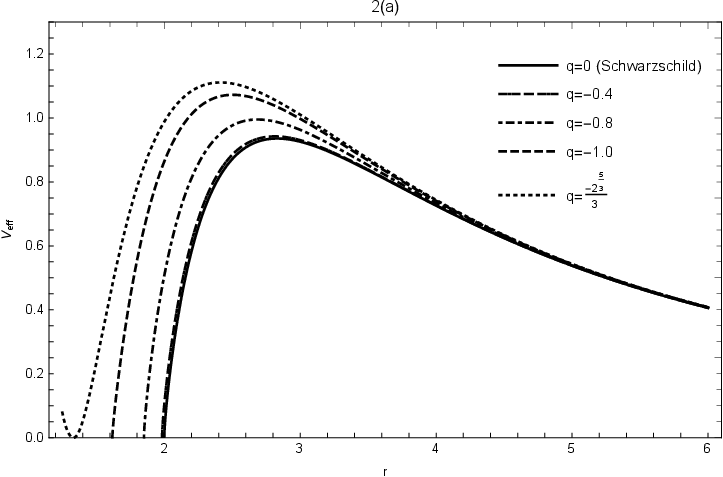}\includegraphics[width=3.3in]{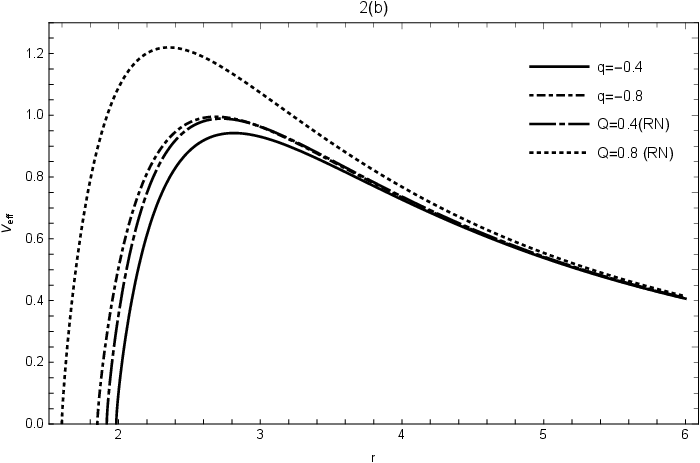}
\end{minipage}
    \caption{The left-hand panel 2(a) shows that the  behavior of the effective potential with different values of $q$ when $\kappa=5$. The left-hand panel 2(b) shows that the comparison of the effective potential in ESTGB-NLED and Reissner-Nordstr$\rm\ddot{o}$m spacetimes when $\kappa=5$.
    }
\label{eff1}
\end{figure}

	Note that Eq.(\ref{E_nled2}) is a Schr$\rm\ddot{o}$dinger-like equation with an effective potential $V_{eff}$. The effective potential is  depicted in Fig.\ref{eff2} when we consider  $\kappa$ as a variable with  the fixed $q=-0.8$. We can observe from Fig.\ref{eff2} that the height of the effective potential barrier becomes larger if $\kappa$ is increased. Furthermore, the location of the peak point moves towards the right with increasing $\kappa$. In addition, we compare the effective potential $V_{eff}$  in Fig.\ref{eff1}(a) with $\kappa$ = 5 under four scenarios, i.e., $q=-0.4$, $q=-0.8$, $q=-1.0$ and $q=-2^{5/3}/3$ respectively. It can be seen that, when the absolute value of the  magnetic charge  is increased, the height of the effective potential barrier  increases and  then the value diminishes and converges to  almost the same value as $r$ increasing. Because this metric is similar to the Reissner-Nordstr$\rm\ddot{o}$m spacetime, we compare the change of the effective potential for the two black hole  when $\kappa$ is set to 5. It is obvious in Fig.\ref{eff1}(b) that  the  maximum value of the effective potential of Reissner-Nordstr$\rm\ddot{o}$m spacetime increases faster with  electric charge than that  of ESTGB-NLED spacetime increases with magnetic charge. This means that the effective potential is more sensitive to the electric charge. Finally, it is worth noting that the effective potential has the form of a single-peak positive definite potential barrier, because it inclines to zero as $r\rightarrow\infty $. In other words, it vanishes at the event horizon  $r_{+}$.

\section{Gray-body factor}
	In this section, we are going to  discuss the grey-body factor for the  Dirac field in 4D ESTGB gravity with a nonlinear electrodynamics, i.e., we calculate the reflection probability and transmission probability. The discussion bases on the six-order WKB method  for the  different  value of the magnetic charge.
	Hawking \cite{Hawking1975} predicted that when the temperature of the black hole is proportional to its surface gravity, the black hole will emit particles, which behaves almost like a black body. So black holes are the thermal systems that has an associated temperature and entropy. Therefore black holes produce radiation when thermodynamic laws is satisfied and take into account the quantum effects \cite{Hawking19761}. Hawking proposed the expression that the evaporation rate of a black hole in a mode with frequency $\omega$ at the event horizon,
	\begin{equation}\label{SSS_ansat11z}
	\Gamma(\omega)=\frac{1}{e^{\beta\omega}\pm 1} \frac{d^{3} k}{(2\pi)^{3}},
	\end{equation}
	where $\beta$ is the inverse of the  black hole temperature, i.e., $\frac{1}{T_{BH}}$, and the plus as well as minus sign are the emission of the  fermions and bosons. However the emission rate, which is measured by the spectator located far away, could be affected by the  geometry situated on the  outside of the event horizon. That is to say, the  geometry situated on the  outside of the event horizon is going to serve as a potential barrier for the radiation that emits from a black hole. The strong gravitational potential near the  event horizon of the  black hole will scatter part of the radiation particles back to the black hole, that is, part of the radiation is reflected back to the black hole. Another part of the particles will pass through the gravitational potential due to the quantum tunneling effect, fly to infinity, and be measured by the remote observer. And  the radiation, which reaches the remote observer through the potential barrier, will no longer  appear as  the form of a black body. Hence we can rewrite the expression of the emission rate recorded by the remote observer with the frequency $\omega$  as,
		\begin{equation}\label{SSS_ansat11z}
	\Gamma(\omega)=\frac{\gamma(\omega)}{e^{\beta\omega}\pm 1} \frac{d^{3} k}{(2\pi)^{3}},
	\end{equation}
    $\gamma(\omega)$ is the gray-body factor related to frequency in the action.

    The gray-body factor is defined as,
    \begin{equation}\label{SSS_ansat11z}
   \gamma(\omega)=\left|\mathcal{T}_{\omega l}\right|^2.
    \end{equation}

   The solution of the second-order differential equation (\ref{Rewritten17}), with the postulation of purely outgoing waves at infinity and purely incoming waves at the event horizon, has the following boundary conditions:
  \begin{eqnarray}\label{E_nled3}
 \psi(r_{\star})\sim\left\{
  \begin{aligned}
  \mathcal{I}_{\omega l} e^{-i\omega r_{\star}} +\mathcal{R}_{\omega l} e^{i\omega
		r_{\star}},~
  r_{\star}~\rightarrow+\infty, \\
  \mathcal{T}_{\omega l }e^{-i\omega r_{\star}},
  ~r_{\star}\rightarrow-\infty .
  \end{aligned}
  \right.
  \end{eqnarray}

 Where $\mathcal{R}_{\omega l}$ and $\mathcal{T}_{\omega l}$ are the reflection coefficient and transmission coefficient, respectively. Due to the conservation of flux,
  $\mathcal{R}_{\omega l}$ and $\mathcal{T}_{\omega l}$ satisfy the following relationship
  \begin{eqnarray}\label{E_nled4}
  \left|\mathcal{R}_{\omega l}\right|^{2}+\left|\mathcal{T}_{\omega l}\right|^2=\left|\mathcal{I}_{\omega l}\right|^{2}.
 \end{eqnarray}

 The phase shift $ \delta_{l} $ can be expressed by

  \begin{eqnarray}\label{E_nled5}
 e^{-2 i\delta_{l}}=(-1)^{l+1} \mathcal{R}_{\omega l}/ \mathcal{I}_{\omega l}.
  \end{eqnarray}	
	
 Now, we discuss the use of WKB method to determine the  gray-body factor \cite{Konoplya2011}. The  gray-body factor relies on the special relation between
  $\omega$ and $V_{m}$, which $V_{m}$ is the peak value of the effective potential $V(r)$. There are three cases with $\omega^{2} \gg V_{m}$,  $\omega^{2} \approx V_{m}$ and  $\omega^{2} \ll V_{m}$ to consider: when $\omega^{2} \gg V_{m}$, i.e.,  the wave will not be reflected back to the black hole by the barrier when the wave with a frequency $\omega $ is higher than the height of the potential barrier. So the transmission  probability $ \mathcal{T}_{\omega l}$ is almost  equal to one.  And the  reflection probability $ \mathcal{R}_{\omega l}$ is close to zero. Almost all of the radiation pass the potential and fly to infinity under this condition. When $\omega^{2} \ll V_{m}$, the transmission  probability $ \mathcal{T}_{\omega l}$ is almost equal to zero as well as the  reflection probability $ \mathcal{R}_{\omega l}$ is close to one. This means that almost all of the radiation is reflected back to the black hole by the  potential. When $\omega^{2} \approx V_{m}$, we compute the gray-body in the limit since the highest precision value is obtained in the WKB approximation.

  Under the WKB approximation, when incident probability $\mathcal{I}_{\omega l}$ is  equal to one the reflection coefficient can be expressed by
  \begin{eqnarray}\label{E_nled44}
  \mathcal{R}_{\omega l}=(1+e^{-2\pi i \alpha})^{-\frac{1}{2}},
  \end{eqnarray}
  \begin{eqnarray}\label{E_nled44}
  \mathcal{T}_{\omega l}=\sqrt{1-\left|(1+e^{-2\pi i \alpha})^{-\frac{1}{2}}\right|^2}.
  \end{eqnarray}
	
  Where $\alpha$ is defined by

  \begin{eqnarray}\label{E_nled44}
  \alpha=\frac{i(\omega^2-V_0)}{\sqrt{-2V_{0}^{(2)}}}-\Lambda_{2}-\Lambda_{3}-\Lambda_{4}-\Lambda_{5}-\Lambda_{6}.
  \end{eqnarray}

   Where $V_{0}$ is the peak value of the effective potential $V(r)$ at $r=r_{0}$. Then,

  \begin{eqnarray}\label{E_nled4w4}
  \begin{aligned}
  &\Lambda_{2}=\frac{1}{\sqrt{-2V^{(2)}_{0}}}[\frac{1}{8}(\frac{V^{(4)}_{0}}{V^{(2)}_{0}})(b^2+\frac{1}{4})-\frac{1}{288}(\frac{V^{(3)}_{0}}{V^{(2)}_{0}})^2(7+60b^2)]\\
   &\Lambda_{3}=\frac{n+\frac{1}{2}}{-2V^{(2)}_{0}}[\frac{5}{6912}(\frac{V^{(3)}_{0}}{V^{(2)}_{0}})^4(77+188b^2)-\frac{1}{384}(\frac{(V^{(3)}_{0})^2V^{(4)}_{0}}{(V^{(2)}_{0})^3})(51+100b^2)\\
   &   +\frac{1}{2304}(\frac{V^{(4)}_{0}}{V^{(2)}_{0}})^2(67+68b^2)-\frac{1}{288}(\frac{V^{(6)}_{0}}{V^{(2)}_{0}})(5+4b^2)+\frac{1}{288}(\frac{V^{(3)}_{0}V^{(5)}_{0}}{(V^{(2)}_{0})^2})(19+28b^2)].
  \end{aligned}
 \end{eqnarray}

   In Eqs. (\ref{E_nled44}) and (\ref{E_nled4w4}), the superscript (2,3,4,5,6) denotes the differentiation with respect to the tortoise coordinate $r_{\star}$ and $b=n+\frac{1}{2}$.  Considering that the expression of $\Lambda_{4}$, $\Lambda_{5}$ and $\Lambda_{6} $ found by \cite{Konoplya2003} is overly cumbersome, we do not describe it in detail here.

\begin{figure}[htbp]
\begin{minipage}[t]{1\linewidth}
\centering
\includegraphics[width=3.3in]{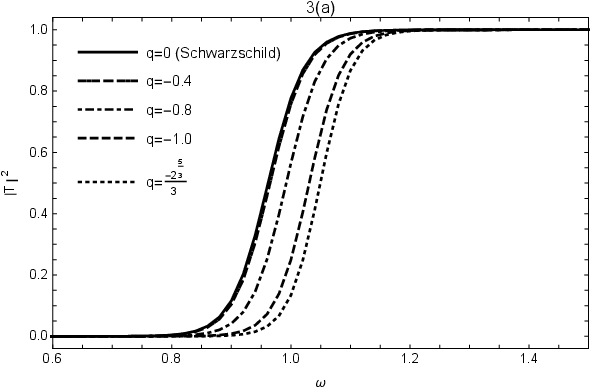}\includegraphics[width=3.3in]{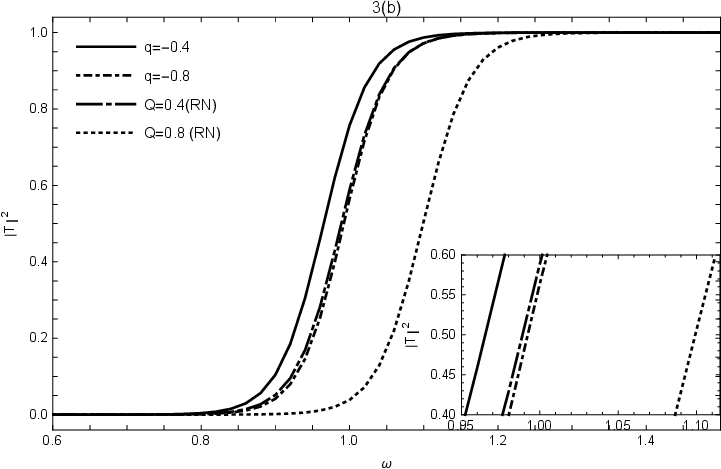}
\end{minipage}
\begin{minipage}[t]{1\linewidth}
\centering
\includegraphics[width=3.3in]{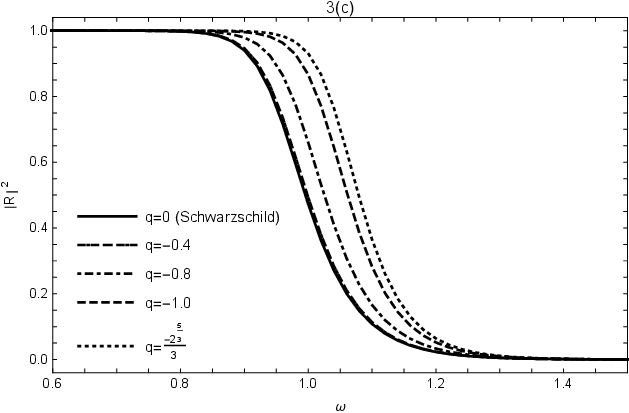}\includegraphics[width=3.3in]{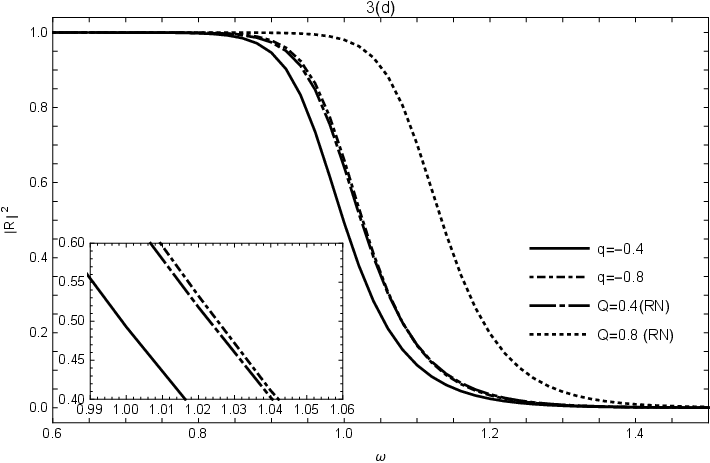}
\end{minipage}
\caption{The top panel 3(a) and 3(b) show the transmission coefficient with $\kappa=5$. The bottom panel 3(c) and 3(d) show the reflection coefficient with $\kappa=5$.}
\label{tran}
\end{figure}

\begin{figure}[htbp]
\begin{minipage}[t]{1\linewidth}
\centering
\includegraphics[width=3.3in]{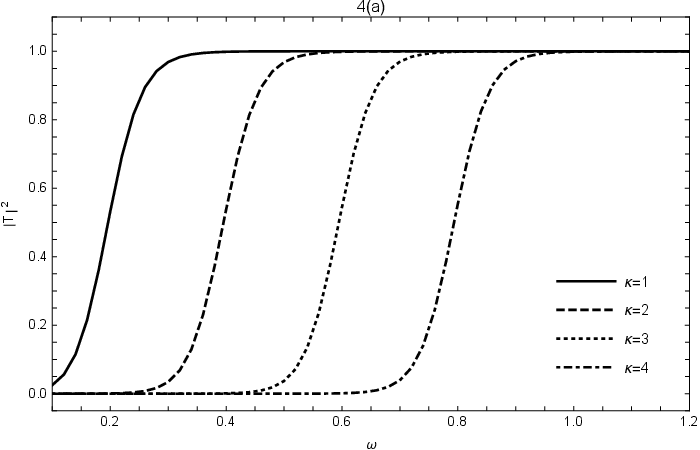}\includegraphics[width=3.3in]{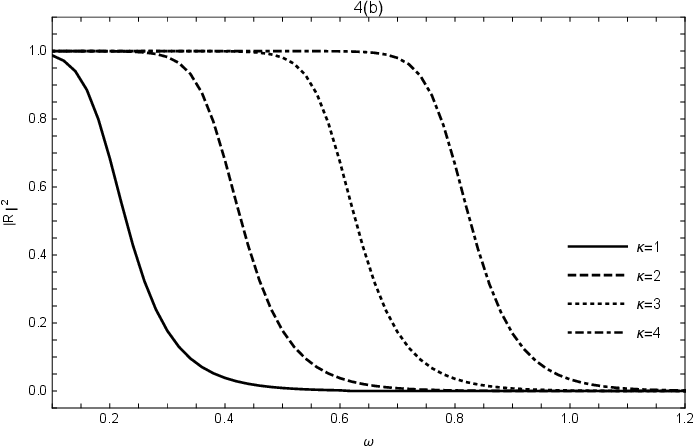}
\end{minipage}
\caption{The left-hand panel 4(a) shows the transmission coefficient with $q=-0.8$. The right-hand panel 4(b) shows the reflection coefficient  with $q=-0.8$.}
\label{tran2}
\end{figure}

  In order to understand the nature of the transmission  and reflection  coefficients, we shall plot them with  the frequency $\omega$ for different values of the magnetic charge. The results for transmission probability are represented in Fig.\ref{tran}(a), where the different values of magnetic charge have been chosen. It can be observed that, for all values of magnetic charge $q$, the transmission coefficient starts from $0$ and reaches $1$ when $\omega$ is increased. Besides, one can see that the transmission coefficient diminishes as the absolute value of the magnetic charge $q$ increases. In other words, the magnetic charge has the behaviour of obstructing the wave from passing through the black hole. The reason may be that the magnetic charge increases the peak value of the effective potential. Fig.\ref{tran}(b) shows the comparison of transmission coefficient for Reissner-Nordstr$\rm\ddot{o}$m  and ESTGB-NLED black hole. The  transmission coefficient of   ESTGB-NLED black hole presents the smaller change than that of Reissner-Nordstr$\rm\ddot{o}$m   black hole when we view the magnetic (electric) charge  as the variable parameter. We also exhibit the reflection coefficient   with $\kappa=5$ in Fig.\ref{tran}(c) and compare it with Reissner-Nordstr$\rm\ddot{o}$m black hole in Fig.\ref{tran}(d). On the contrary, the reflection coefficient starts from $1$ and reaches $0$ with the increase of $\omega$. Furthermore, as shown in Fig.\ref{tran2}, we present the results of the transmission and reflection coefficients in the case where $q=-0.8$ and $\kappa$ varies from $1$ to $4$. When $\kappa$ is increasing, the reflection coefficient  becomes larger and the transmission coefficient  becomes smaller. The reason for this behavior is  that  the peak value of the effective potential increases with the increase of $\kappa$.

\section{Absorption cross section}

In this section, we calculate the absorption cross section for the Dirac field, which is defined as the ratio of the number of particles absorbed by the black hole to the incident particle flux. Benone \cite{Benone104053} proposed the partial wave method to get total absorption cross section as follows,
	\begin{equation}\label{SSS_ansatz}
   \sigma_{abs}=\sum_{l=0}^{\infty}\sigma_{abs}^{l},
   \end{equation}
	and the partial absorption cross section is given by
	
	\begin{equation}\label{SSS_ansatz}
	\sigma_{abs}^{l}=\frac{\pi}{\omega^{2}}(2l+1)(1-\left|e^{-2 i\delta_{l}}\right|^2),
	\end{equation}
   we substitute the phase shift $e^{-2 i\delta_{l}}$ of different $l$ subwaves into the Eq.(33), then we can get another expression,
	\begin{equation}\label{SSS_ansatz}
   \sigma_{abs}=\frac{\pi}{\omega^{2}}\sum_{l=0}^{\infty}(2l+1)(1-{\left|\mathcal{R}_{\omega l}\right|^2}).
   \end{equation}	

\begin{figure}[htbp]
	\centering
	{
		\begin{minipage}[t]{0.5\linewidth}
			\centering
			\includegraphics[width=3.3in]{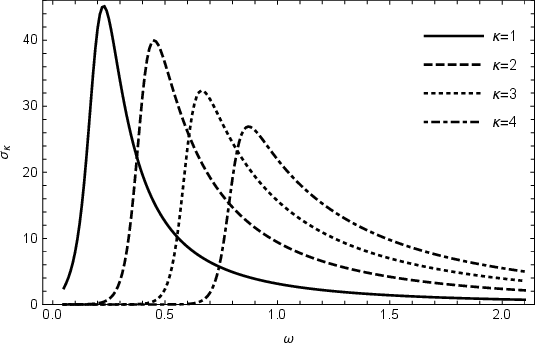}
		\end{minipage}%
	}%
		\caption{Behavior of  the partial absorption cross section with different values of $\kappa$ at $q=-0.8$.}
\label{sec1}
\end{figure}

\begin{figure}[htbp]
\begin{minipage}[t]{1\linewidth}
\centering
\includegraphics[width=3.3in]{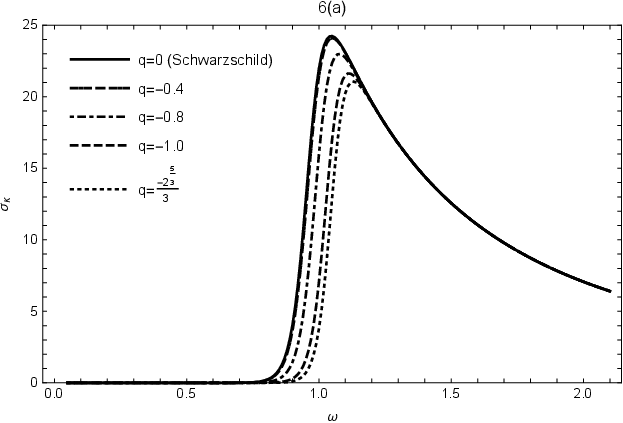}\includegraphics[width=3.3in]{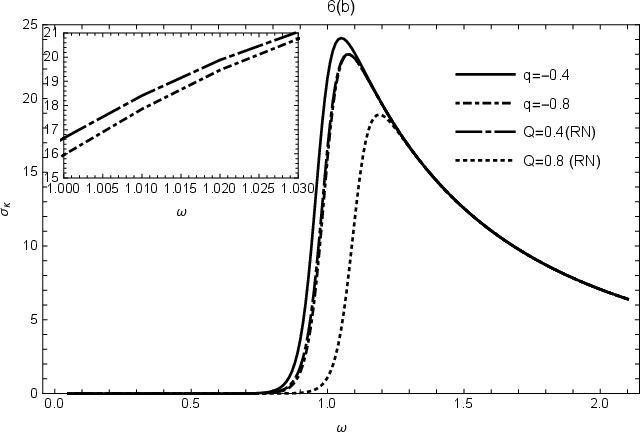}
\end{minipage}
\caption{The left panel 6(a) shows behavior of the total partial absorption cross section for the different value of magnetic charge when $\kappa=5$.
         The right panel 6(b) shows behavior of the total partial absorption cross section of Reissner-Nordstr$\rm\ddot{o}$m black hole when $\kappa=5$, for comparison.}
\label{sec2}
\end{figure}

\begin{figure}[htbp]
\begin{minipage}[t]{1\linewidth}
\centering
\includegraphics[width=3.3in]{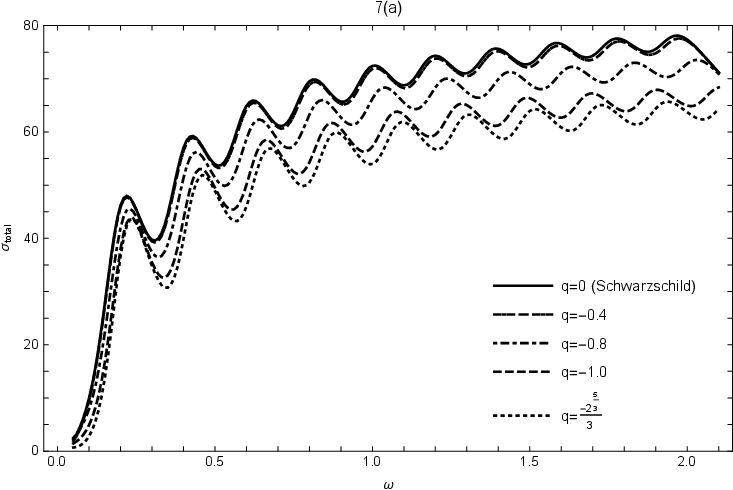}\includegraphics[width=3.3in]{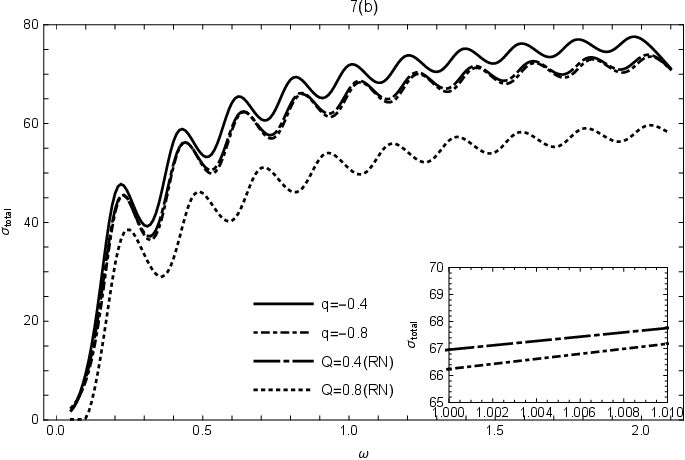}
\end{minipage}
\caption{The left panel 7(a)  shows behavior of absorption cross section from $\kappa=1$ to $\kappa=10$ for the different value of magnetic charge.
         The right panel 7(b)  shows behavior of absorption cross section of Reissner-Nordstr$\rm\ddot{o}$m black hole for comparison.
}
\label{tal}
\end{figure}

	Considering the effects of the Dirac field, the results of the partial absorption cross section are shown in Fig.\ref{sec1} and Fig.\ref{sec2}. One can see from Fig.\ref{sec1} that the peak value of the partial absorption cross section decreases and the location of the maximum moves to the right when $\kappa$ is increased. By observing the curves in Fig.\ref{sec2}(a) for different magnetic charge $q$, one can obtain that the partial absorption cross section tends to almost the same value on the low and high frequencies, and the peak value moves slightly to the right when the absolute value of $q$ is increased. In Fig.\ref{sec2}(b), we compare the partial absorption cross section of Reissner-Nordstr$\rm\ddot{o}$m and ESTGB-NLED black holes.  We note that the partial absorption cross section of the two types black hole also goes  to almost the same value on the low- and high-frequency regions due to the effective potential has the form of a single-peak positive. The effective potential is more sensitive to the electric charge. Specifically, the electric charge can hinder the passage of the wave more. So  in the mid-frequency the curve change of the partial absorption cross section of the ESTGB-NLED black hole is not as obvious as that of the Reissner-Nordstr$\rm\ddot{o}$m black hole  when we take the  magnetic (electric) charge as the variable.
 As shown in Fig.\ref{tal}(a), we draw the results of the total absorption cross section from $\kappa=1 $ to $\kappa=10$ with the magnetic charge $q=-0.4, q=-0.8, q=-1.0$ and $q=-2^{5/3}/3$ respectively. We can see that the total absorption cross section increases and then tends to a stable value when we increase the  frequency $\omega$. We also obtain that, as we increase the  absolute value of the magnetic charge, the total absorption cross section  gradually diminishes. That is to say that, the magnetic charge weakens the absorption for the Dirac field. This is in agreement with the results presented in Ref. \cite{Huang2015}. Additionally, we plot the total absorption cross section for the Reissner-Nordstr$\rm\ddot{o}$m black hole  in Fig.\ref{tal}(b), for comparison purposes.
Compared  with the effective potential of Reissner-Nordstr$\rm\ddot{o}$m black hole, we find that the effective potential of the ESTGB-NLED black hole changes slower.  We can say that the  magnetic charge has a smaller effect on the total absorption cross-section of the Dirac field than the electric charge. Therefore, the variation of  total absorption cross section of the ESTGB-NLED black hole with the magnetic charge is not as pronounced as for  the Reissner-Nordstr$\rm\ddot{o}$m black hole black hole with the charge.

\section{Conclusions}

In the preceding sections, we have studied the gray-body factor and the absorption cross section for the massless Dirac field of the black hole, which is the solution of the 4D ESTGB gravity in the context of the nonlinear electrodynamics. Due to the fact that the solution is characterized by the ADM mass $m$ and the magnetic charge $q$,  the black holes will have different structure according to the different choices of these parameters. Therefore, in order to consider  the generality, we have studied the case that the black hole is non-extreme where $m=1$ and $-2^{(\frac{5}{3})}/3 < q < 0 $, which is similar to the Reissner-Nordstr$\rm\ddot{o}$m spacetime. Specifically, we have plotted the effective potentials in Fig.1 and Fig.2 for two cases. We have found that the effective potential of the Dirac field  is more sensitive  for the electric charge owing to that  the variation of the effective potential of Reissner-Nordstr$\rm\ddot{o}$m spacetime is more obvious than that of ESTGB-NLED spacetime. Besides, we have carried out numerical calculations to get the gray-body factors using the sixth-order  WKB approximations.  We have shown the changes of the transmission and reflection coefficients with respect to the magnetic charge in Fig.3, respectively.  We have observed that due to the magnetic charge  the reflection coefficient is increasing and the transmission coefficient is decreasing comparing the Schwarzschild spacetime. In other words, we have obtained that the magnetic charge impedes the wave from passing the black hole. Moreover, when $\kappa$ is increasing, the reflection coefficient becomes larger and the transmission coefficient becomes smaller. It has been discovered in Fig.\ref{tal}(a) that the total absorption cross section of the Dirac field  decreases when we augment the absolute value of the magnetic charge, but increases with the  increasing of frequency. Finally, in Fig.\ref{tal}(b), we have compared the total absorption cross section  for the Reissner-Nordstr$\rm\ddot{o}$m and ESTGB-NLED black hole. It has been found that  the absorption cross section of the  Dirac field is more sensitive to  electric charge than magnetic charge in these two types of black hole.

\section*{Declaration of Competing Interest}
The authors declare that they have no known competing financial interests or personal relationships that could have appeared to influence the work reported in this paper.

\section*{Acknowledgments}
This work was supported partly by the National Natural Science Foundation of China (Grants No. 12065012, No. 12065013), Yunnan High-level Talent Training Support Plan Young \& Elite Talents Project (YNWR-QNBJ-2018-360) and the Fund for Reserve Talents of Young and Middle-aged Academic and Technical Leaders of Yunnan Province (Grant No. 2018HB006).


\begin{thebibliography}{99}
		
\bibitem{1}Akiyama,~K.~et al.:
Astrophys. J. Lett. \textbf{875}(2019)L1.

\bibitem{Riess1998}
Riess,~A.G.~et al.:
Astron. J. \textbf{116}(1998)1009-1038.

\bibitem{Fradkin1985}
Fradkin,~E.S. and Tseytlin,~A.A.:
Nucl. Phys. B \textbf{261}(1985)1-27.

\bibitem{Capozziello2007}
Capozziello,~F.~et al.:
Mon. Not. Roy. Astron. Soc. \textbf{375}(2007)1423-1440.

\bibitem{Doneva2018}
Doneva,~D.D.~et al.:
Phys. Rev. D \textbf{98}(2018)104056.

\bibitem{Woodard02210}
Motohashi,~H. and Suyama,~T.:
JHEP \textbf{09}(2020)032.

\bibitem{Doneva2019}
Doneva,~D.D.~et al.:
Phys. Rev. D \textbf{99}(2019)104045.

\bibitem{Kanti1996}
Kanti,~P.~et al.:
 Phys. Rev. D \textbf{54}(1996)5049-5058.

\bibitem{scalariz_Doneva}
Doneva,~D.D.~et al.:
Phys. Rev. D \textbf{102}(2020)064042.

\bibitem{Canate2020}
Ca\~nate,~P. and Perez Bergliaffa,~S.E.:
Phys. Rev. D \textbf{102}(2020)104038.

\bibitem{Macedo2014}
Macedo,~C.F.B. and Crispino,~L.C.B.:
Phys. Rev. D \textbf{90}(2014)064001.

\bibitem{Hawking1976}
Hawking,~S.W.:
Phys. Rev. D \textbf{14}(1976)2460-2473.

\bibitem{Kanti2002}
Kanti,~P. and March-Russell,~J.:
Phys. Rev. D \textbf{66}(2002)024023.

\bibitem{Songbai2010}
Chen,~S. and Jing,~J.:
Phys. Lett. B \textbf{691}(2010)254-260.

\bibitem{Ama-Tul-Mughani2021}
Ama-Tul-Mughani,~Q.~et al.:
Astropart. Phys. \textbf{132}(2021)102623.

\bibitem{Sharif2020}
Sharif,~M. and Ama-Tul-Mughani,~Q.:
Phys. Dark Univ. \textbf{27}(2020)100436.

\bibitem{Qanitah Ama-Tul-Mughani2020}
Sharif,~M. and Ama-Tul-Mughani,~Q.:
PTEP \textbf{2020}(2020)033E01.

\bibitem{Konoplya2021}
Konoplya,~R.A.:
Phys. Lett. B \textbf{823}(2021)136734.

\bibitem{Cvetic1998}
Cvetic,~M. and Larsen,~ F.:
Phys. Rev. D \textbf{57}(1998)6297-6310.

\bibitem{Zhang2018}
Zhang,~C.Y., Li,~P.C. and Chen,~B.:
Phys. Rev. D \textbf{97}(2018)044013.

\bibitem{Miao2017}
Miao,~Y.G. and Xu,~Z.M.:
Phys. Lett. B \textbf{772}(2017)542-546.

\bibitem{Konoplya2011}
Konoplya,~R.A. and Zhidenko,~A.:
Phys. Lett. B \textbf{686}(2010)199-206.

\bibitem{Blome1984}
Blome,~H.J., and Mashhoon,~B.:
Phys. Lett. A \textbf{100}(1984)231-234.

\bibitem{Schutz1985}
Schutz,~B.F. and Will,~C.M.:
Astrophys. J. Lett. \textbf{291}(1985)L33-L36.

\bibitem{Iyer1987}
Iyer,~S. and Will,~C.M.:
Phys. Rev. D \textbf{35}(1987)3621.

\bibitem{Konoplya2003}
Konoplya,~R.A.:
Phys. Rev. D \textbf{68}(2003)024018.

\bibitem{Matyjasek2017}
Matyjasek,~J. and Opala,~M.:
Phys. Rev. D \textbf{96}(2017)024011.

\bibitem{Wongjun2020}
Wongjun,~P.~et al.:
Phys. Rev. D \textbf{101}(2020)124033.

\bibitem{Cai2020}
Cai,~X.C. and Miao,~Y.G.:
Phys. Rev. D \textbf{101}(2020)104023.

\bibitem{Hu2020}
Hu,~Y.~et al.:
EPL \textbf{128}(2019)50006.

\bibitem{Liang2018}
Liang,~J.:
Commun. Theor. Phys. \textbf{70}(2018)695.

\bibitem{Saleh2018}
Saleh,~M.,Thomas,~B.B. and Kofane,~T.C.:
Eur. Phys. J. C \textbf{78}(2018)325.

\bibitem{Liang20018}
Liang,~J.:
Chin. Phys. Lett. \textbf{35}(2018)050401.

\bibitem{Aragon2021}
Arag\'on,~A.~et al.:
Phys. Rev. D \textbf{103}(2021)064006.

\bibitem{Li2017}
Li,~J., Lin,~K. and Wen,~H.:
Adv. High Energy Phys. \textbf{2017}(2017)5234214.

\bibitem{Wang2021}
Wang,~M.~et al.:
Eur. Phys. J. C \textbf{81}(2021)469.

\bibitem{Jawad2020}
Jawad,~A.~et al.:
Mod. Phys. Lett. A \textbf{35}(2020)2050298.

\bibitem{Sharif2021}
Sharif,~M. and Khan,~A.:
[arXiv:2109.06010 [gr-qc]].

\bibitem{Guo2013}
Guo,~G.:
Eur. Phys. J. C \textbf{73}(2013)2573.

\bibitem{Futterman}
Futterman,~J.A.H.~et al.:
1988 (Cambridge: Cambridge Uni-versity Press) p. 254.

\bibitem{Sanchez1978}
Sanchez,~N.G.:
Phys. Rev. D \textbf{18}(1978)1030.

\bibitem{Unruh1976}
Unruh,~W.G.:
Phys. Rev. D \textbf{14}(1976)3251-3259.

\bibitem{Crispino2007}
Crispino,~L.C.B.~et al.:
Phys. Rev. D \textbf{75}(2007)104012.

\bibitem{Jung2004}
Jung,~E.~et al.:
Phys. Lett. B \textbf{602}(2004)105-111.

\bibitem{Crispino2008}
Crispino,~L.C.B. and Oliveira,~E.S.:
Phys. Rev. D \textbf{78}(2008)024011.

\bibitem{Songbai Chen2014}
Liao,~H.~et al.:
Phys. Lett. B \textbf{728}(2014)457-461.

\bibitem{Macedo2013}
Macedo,~C.F.B.~et al.:
Phys. Rev. D \textbf{88}(2013)064033.

\bibitem{Leite2017}
Leite,~L.C.S.~et al.:
Phys. Lett. B \textbf{774}(2017)130-134.

\bibitem{Huang2019}
Huang,~H.~et al.:
Gen. Rel. Grav. \textbf{51}(2019)22.

\bibitem{Huang2015}
Huang,~H.~et al.:
Gen. Rel. Grav. \textbf{47}(2015)8.

\bibitem{2018}
Leite,~L.C.S.~et al.:
Phys. Rev. D \textbf{98}(2018)024046.

\bibitem{Anacleto2020}
Anacleto,~M.A.~et al.:
Phys. Lett. B \textbf{803}(2020)135334.

\bibitem{Magalhaes2020}
Magalh\~aes,~R.B.~et al.:
Eur. Phys. J. C \textbf{80}(2020)386.

\bibitem{Junior2020}
Lima,~H.C.D.~et al.:
Phys. Lett. B \textbf{811}(2020)135921.

\bibitem{Benone2018}
Benone,~C.L.~et al.:
Int. J. Mod. Phys. D \textbf{27}(2018)1843012.

\bibitem{Brill1957}
Brill,~D.R. and Wheeler,~J.A.:
Rev. Mod. Phys. \textbf{29}(1957)465-479.

\bibitem{Cho2005}
Cho,~H.T. and Lin,~Y.C.:
Class. Quant. Grav. \textbf{22}(2005)775-790.

\bibitem{Cooper1995}
Cooper,~F.~et al.:
Phys. Rept. \textbf{251}(1995) 267-385.

\bibitem{Hawking1975}
Hawking,~S.W.:
Commun. Math. Phys. \textbf{43}(1975)199-220.

\bibitem{Hawking19761}
Hawking,~S.W.:
Phys. Rev. D \textbf{13}(1976)191-197.

\bibitem{Benone104053}
Benone,~C.L.~et al.:
Phys. Rev. D \textbf{89}(2014)104053.
	
\end{thebibliography}
\end{document}